\documentclass[a4paper]{article}

\usepackage{INTERSPEECH2019}

\graphicspath{{figures/}}

\renewcommand{\vec}[1]{\boldsymbol{\mathrm{#1}}}
\newcommand{\mtx}[1]{\boldsymbol{\mathrm{#1}}}

\newcommand*\dif{\mathop{}\!\mathrm{d}}

\usepackage{algorithm,algorithmic} 
\usepackage[norelsize,algo2e]{algorithm2e} 
\SetKwInput{KwInput}{Input} 

\newcommand\E{\mathbb{E}}

\newcommand{\diag}[1]{\ensuremath{ \mathsf{diag} \left(#1\right)}}

\renewcommand{\vec}{\boldsymbol} 

\usepackage{xcolor}
\usepackage{enumitem}

\title{Extrapolating False Alarm Rates in Automatic Speaker Verification}

\name{Alexey Sholokhov$^{1}$, Tomi Kinnunen$^2$, Ville Vestman$^2$, Kong Aik Lee$^3$}
\address{
  $^1$Huawei Technologies Ltd., Moscow, Russia\\
  $^2$Computational Speech Group, University of Eastern Finland, Finland\\
  $^3$Biometrics Research Laboratories, NEC Corporation, Tokyo, Japan}
\email{sholokhov.alexey@huawei.com, tkinnu@cs.uef.fi, vvestman@cs.uef.fi, kongaik.lee@nec.com}

\begin{document}

\maketitle

\setlength{\abovedisplayskip}{2pt}
\setlength{\belowdisplayskip}{2pt}

\addtolength{\parskip}{-0.5mm}

\begin{abstract}
Automatic speaker verification (ASV) vendors and corpus providers would both benefit from tools to reliably extrapolate  performance metrics for large speaker populations \emph{without collecting new speakers}. We address false alarm rate extrapolation under a worst-case model whereby an adversary identifies the closest impostor for a given target speaker from a large population. Our models are generative and allow sampling new speakers. The models are formulated in the ASV detection score space to facilitate analysis of arbitrary ASV systems.
\end{abstract}

%

\section{Introduction}

\emph{How unique the human voice is?} This question is clearly relevant for practical deployment of automatic speaker verification (ASV) technology --- yet, scarcely addressed \cite{Nautsch2015} due to the open-ended nature of the question. Unlike passwords that have zero uncertainty conditioned on person's identity \cite{Takahashi2014_biometric}, the human voice is subject to both extrinsic and intrinsic variations, none of which are deterministic. `Uniqueness' thus depends both on data conditions \emph{and} the observer (\emph{e.g.} a specific ASV system or listener). In our recent work \cite{Sholokhov2020-CSL}, we addressed an alternative, more tangible question:
    \begin{quote}
        \emph{Given a specific ASV system (black-box) and evaluation corpus, how does the false alarm rate behave with an increased number of speakers?}
    \end{quote}
To be precise, we modeled the sampling process of nontarget detection scores of a given ASV system through a probabilistic generative model to enable indefinite increasing of the impostor population size without having to collect new speech data. The assumption is that the underlying sampling process, governed by the properties of the ASV system (treated as a black-box) and corpus, remains fixed. Drawing a random nontarget score proceeds in two steps. 
First, we draw a \emph{random pair of speakers} implicitly represented by a Gaussian distribution which models similarity scores between these two speakers. Second, we draw a \emph{random score} from that distribution.


In \cite{Sholokhov2020-CSL} we also revised the notion of `nontarget speaker'. Apart from efforts devoted to the study of spoofing attacks \cite{Wu2017-ASVspoof}, standard evaluation benchmarks of ASV technology \cite{Greenberg2020-two-decades} assume nontarget speakers to be \emph{non-proactive} or \emph{zero-effort} impostors --- other \emph{random} speakers paired up with targets. We, instead, considered \emph{worst-case} impostors with a \emph{deterministic}, proactive imposture policy: given a target speaker of interest (for instance, a notable politician), the adversary identifies the closest impostor to the given target from a large population (such as the Internet) to increase the chance of this impostor to be accepted as the targeted speaker. This
is an instance of \emph{adversarial attack} \cite{Szegedy2014-intriguing,Biggio2018-adversarial} on ASV \cite{Kreuk2018-fooling,das2020attackers}. The general motivations are to identify loopholes of ASV and to develop defense mechanisms against them. 

In this study we 
improve upon the generative model presented in \cite{Sholokhov2020-CSL}. Despite demonstrating expected overall trends, the predicted false alarm rates were substantially overestimated, particularly at high ASV thresholds (proxies of high-security applications). To tackle this shortcoming, we propose a discriminative training method which uses empirical estimates of false alarm rates as targets.
The setup is similar to standard regression tasks except that our primary goal is \emph{extrapolation} --- making predictions substantially beyond the range of inputs in the training set. In our context, this means predicting false alarm rate of an ASV system for a population of, say, $100,000$ nontarget speakers but with access to data from only $1000$ speakers. Without additional assumptions on the predictor functions, standard regression methods available in machine learning libraries have higher risk of producing meaningless results (see \cite{Martius2017-extrapolation}). 

In general, the task of learning \emph{interpretable} functional dependencies has received far less attention within machine learning compared to natural sciences, where discovering \emph{physically plausible} models is important. To obtain more trusthworthy predictions, we build upon a regressor which takes into account the specifics of detection score distribution governed by unobserved similarities between speakers.
Specifically, it uses a generative model of ASV scores together with an estimator of false alarm rates in 
a single prediction pipeline. 


Another novelty of this work is modeling the generation of nontarget scores using \emph{probabilistic linear discriminant analysis} (PLDA) \textbf{in detection score space}. 
PLDA \cite{PrinceLFME12} --- a generative model in the space of vector representations of speech utterances (\emph{e.g.} i-vectors or x-vectors) --- is well-known by ASV researchers. Our formulation, however, differs substantially from this familiar use case as our modeling takes place in the detection score, rather than vector space.
We use PLDA to generate `new' detection scores. The scores used for training can, but are not required to be outcomes of trial comparisons by an actual PLDA model.  
We learn PLDA whose log-likelihood ratio scores are designed to approximate distribution of detection scores of \emph{any} ASV system. 
Similar to \cite{Sholokhov2020-CSL}, our models 
require no other data than ASV scores (and their labels). In specific, we do \emph{not} need any speaker embeddings to train our PLDA score generator.







\section{Preliminaries}

We begin with a brief review of some necessary technical background on false alarm rate, its extrapolation, and PLDA.

\subsection{False alarm rate}
The \emph{false alarm} (FA) rate is defined as 
\begin{equation}\label{eq:false-alarm-rate}
\mathrm{P}_\text{FA}(\tau) \equiv \int_\tau^{\infty} p(s|\text{non}) \dif s,
\end{equation}
where $\tau \in \mathbb{R}$ is a detection threshold and $p(s|\text{non})$ is the probability density of nontarget scores of an ASV system. The FA rate can be written as the expectation $\E_{s \sim p(s|\theta_\text{non})}[\mathbb{I}\{s > \tau\}]$, and 
approximated by Monte-Carlo (MC) sampling as
\begin{equation} \label{eq:fa-mc}
\mathrm{P}_\text{FA}(\tau) \approx \frac{1}{R} \sum_{r=1}^R \mathbb{I}\{s_r > \tau\}, \;\; s_r \sim p(s|\theta_\text{non}),
\end{equation}
where $r=1,\dots,R$ are the indices of the nontarget trials and $\mathbb{I}\{\cdot\}$ is an indicator function. Each nontarget trial consists of a pairwise comparison of utterances from two different speakers (conversely, a target trial constitutes a pairwise comparison of utterances from the same speaker). In the special case when every unique speaker pair in a trial list has the same number of trials, $L$, the above estimator is the same as averaging \emph{speaker-pair specific} FA rates:
\begin{equation} \label{eq:fa-mc-nested}
\frac{1}{R} \sum_{r=1}^R \mathbb{I}\{s_r > \tau\} = \frac{1}{T} \sum_{i=1}^T 
\frac{1}{L} \sum_{\ell=1}^{L} \mathbb{I}\{s_{i,\ell} > \tau\}
\end{equation}
where $s_{i,\ell}$ denotes the $\ell^\text{th}$ score from the $i^\text{th}$ speaker pair, $T$ is the number of unique speaker pairs and $R=T\cdot L$. 

This reformulation of \eqref{eq:fa-mc} leads to an alternative estimator of $P_\text{FA}$ as presented in \cite{Sholokhov2020-CSL}:
\begin{equation} \label{eq:fa-mc-general}
\mathrm{P}_\text{FA}(\tau) \approx \frac{1}{T} \sum_{i=1}^T P_\text{FA}^{(i)}(\tau),  P_\text{FA}^{(i)}(\tau) = \frac{1}{|\mathcal{S}_i|} \sum_{s_\ell \in \mathcal{S}_i} \mathbb{I}\{s_{\ell} > \tau\},
\end{equation}
where $\mathcal{S}_i$ is the set of scores for the $i^\text{th}$ speaker pair consisting of an enrolled (target) speaker and an impostor selected randomly from a dataset and $P_\text{FA}^{(i)}(\tau)$ is the corresponding speaker-pair specific FA rate. The following discussion is based on the fact that selecting a random impostor is equivalent to selecting a random subset of $N$ speakers, followed by selecting a random speaker from this subset.
Thus, \eqref{eq:fa-mc-general} can be interpreted as averaging of results of $T$ stochastic simulations, where both the target speaker and the impostor subset are randomly drawn from a given database. This is in line with typical ASV trial designs, where (zero-effort) impostors can be considered as random speakers with different identity. 

This view of \eqref{eq:fa-mc-general} allows us to consider several alternative policies to choose an impostor. 
We consider \emph{worst-case} impostor 
that is the closest match to a given target speaker. The adversary might locate the closest impostor using a speaker identification system \cite{Vestman2020_mimicry_attack} or by other means. In \cite{Sholokhov2020-CSL} we proposed a new metric, \emph{worst-case FA rate with $N$ impostors}, abbreviated $\mathrm{P}^N_\text{FA}(\tau)$. It represents a scenario where target speakers are scored against their closest impostors. We also introduced generative model of nontarget scores to allow $N$ to exceed the number of speakers in the corpus. This allows extrapolation of FA rates for arbitrarily-sized impostor population.

\subsection{Probabilistic linear discriminant analysis (PLDA)}

PLDA \cite{PrinceLFME12} models between- and within-class distributions of high-dimensional vectors using low-dimensional subspaces. In ASV, PLDA is used to model distributions of \emph{speaker embeddings} and for same/different speaker hypothesis testing. PLDA was revised in \cite{Kenny-2010} and \cite{Brummer2010-two-cov} (see also \cite{Sizov2014-unifying}). We use the so-called \emph{two-covariance} PLDA \cite{Brummer2010-two-cov}. It models the $j$th embedding of the $i$th speaker by
    \begin{equation}\label{eq:2cov}
        \vec{\phi}_{i,j} = \vec{b} + \vec{y}_i + \vec{\varepsilon}_{i,j},
    \end{equation}
where $\vec{b} \in \mathbb{R}^D$ is the center of the embedding space, $\vec{y}_i \in \mathbb{R}^D$ is a latent \emph{speaker identity variable} with normal prior $\mathcal{N}(\vec{0},\mtx{B})$, and $\vec{\varepsilon}_{i,j} \in \mathbb{R}^D$ is \emph{residual} with prior $\mathcal{N}(\vec{0},\mtx{W})$. $\mtx{B}$ and $\mtx{W}$ are the \emph{between-} and \emph{within-}class covariance matrices. 
%
The parameters $\vec{\theta}_\text{plda}=\{\vec{b},\mtx{B},\mtx{W}\}$ are typically estimated via the \emph{expectation-maximization} (EM) algorithm \cite{Dempster77maximumlikelihood,Bishop-book} using a set of development speakers (different from target speakers).

At the recognition stage, $\vec{\theta}_\text{plda}$ is used for computing \emph{log-likelihood ratio} (LLR) score for a given pair of enrollment and test utterances, as 
	\begin{equation}\label{eq:plda-llr-score}
    	s(\vec{\phi}_\text{e},\vec{\phi}_\text{t}) = \log \frac{p(\vec{\phi}_\text{e},\vec{\phi}_\text{t}|H_0,\vec{\theta}_\text{plda})}{p(\vec{\phi}_\text{e},\vec{\phi}_\text{t}|H_1,\vec{\theta}_\text{plda})},
    \end{equation}
where 
$H_0$ and $H_1$ denote, respectively, the target (same speaker) and nontarget (different speaker) hypotheses. $H_0$ assumes that $\vec{\phi}_\text{e}$ and $\vec{\phi}_\text{t}$ share the same latent identity variable and $H_1$ assumes that their latent identity variables are different. The score \eqref{eq:plda-llr-score} is given by a closed-form expression --- see \cite{Rohdin-2014}. 


\subsection{PLDA in the score space}

In this work, we do \emph{not} use PLDA to model speaker embeddings. For generality, all our modeling takes place in the detection scores space. We use PLDA to model the distribution of empirical scores of \emph{any} ASV system --- whether or not based on a PLDA back-end. Note, first, that  \eqref{eq:plda-llr-score} represents a deterministic function
$s\colon\mathbb{R}^{2D}\rightarrow \mathbb{R}$ that assigns a real number to any pair of embeddings. Concerning performance assessment, the embeddings are not relevant. The distribution of the detection scores (rather, the \emph{order} of the scores) is a complete description of the detection error trade-off (DET) behavior of a given system \cite{Brummer2010-PhD}. Second, note that PLDA is a generative model --- it allows sampling new `speakers' in the $\vec{y}$-space. We want to fit a PLDA model whose score generation mechanism produces distributions similar to the given empirical scores. 

To this end, we first note that PLDA is heavily over-parameterized from the perspective of LLR score order preservation. 
A centered PLDA model ($\vec{b} = \vec{0}$) uses $D(D+1)/2$ parameters for each of the matrices $\mtx{B}$ and $\mtx{W}$, totaling $D^2 + D$ \cite{Sizov2014-unifying}.
In fact, we need only $D$ numbers.
Note that any invertible linear transformation of the feature space leaves the order of scores unchanged. Hence, it does not alter a DET-curve.
We can therefore perform \emph{simultaneous diagonalization} \cite{Fukunaga-book, Wang2017-joint} of the within-class and between-class covariance matrices such that (i) $\mtx{B}$ becomes an identity matrix and (ii) $\mtx{W}$ becomes diagonal: $\mtx{W} = \diag{d_1, \dots, d_D}$. 
Therefore, a PLDA model can be defined through $D$  nonnegative numbers. We use this minimal parametrization in our experiments.

\section{Extrapolating false alarm rates} \label{sec:fa-extrapolation}

With the above preliminaries, we are now set to present models to produce predictions of $\mathrm{P}^N_\text{FA}(\tau)$. We consider two different types of models. Our previous model \cite{Sholokhov2020-CSL} is a special case of a \textbf{location-scale} model described below, while \textbf{PLDA-based} model is a new proposal. Both models serve to approximate the distribution of \emph{sets of scores} between a random target speaker and the closest impostor selected from a random set of $N$ impostors. These sets of scores can be viewed as outcomes of the generative process in Algorithm \ref{algo:scores-n-best}. 
\begin{algorithm}[!h] 
\setlength{\abovedisplayskip}{1pt}
\setlength{\belowdisplayskip}{1pt}
\SetAlgoLined
 \For{$i = 1,\dots,T$}
 {
  Sample random enrolled (target) speaker, $\vec{y}_\text{e}^{(i)}$.\\
  Sample $N$ random test speakers, $\vec{y}_{\text{t},1}^{(i)}, \vec{y}_{\text{t},2}^{(i)},\dots,\vec{y}_{\text{t},N}^{(i)}$. \\
  Find the closest speaker $\vec{y}_{\text{t},k}^{(i)}$, where\\ 
  $k = \arg\max_j\,\, \mathsf{sim} (\vec{y}_{e}^{(i)}, \vec{y}_{\text{t},j}^{(i)})$ \\
  Sample scores $\mathcal{S}_i=\{s_\ell\}_{\ell=1}^{L_i}$ between $\vec{y}_\text{e}^{(i)}$ and $\vec{y}_{\text{t},k}^{(i)}$.
 }
\caption{} \label{algo:scores-n-best}
\end{algorithm}


Here, $\mathsf{sim}(\cdot, \cdot)$ is any speaker similarity measure. Since explicit speaker representation are not available in the general case the similarity function has to be computed from a set of speaker-pair specific scores. This case includes estimating $\mathrm{P}^N_\text{FA}(\tau)$ from empirical scores.
We use the mean value of scores as a similarity measure. Given a sampled \emph{set} of score sets $\{\mathcal{S}_1, \dots, \mathcal{S}_T\}$, we compute the corresponding MC estimates of the speaker-pair conditioned FA rates $\{\mathrm{P}_\text{FA}^{(1
)}(\tau), \dots, \mathrm{P}_\text{FA}^{(T)}(\tau)\}$ --- the individual terms of the sum in \eqref{eq:fa-mc-general}. Averaging them yields an estimate of $\mathrm{P}^N_\text{FA}(\tau)$. We now
describe two generative models that allow sampling scores according to Algorithm \ref{algo:scores-n-best} for an arbitrary $N$. Each model can be trained on sets of speaker-pair specific ASV scores and further be used for FA rate extrapolation.


\subsection{Location-scale models}

Our first family of models assumes the distribution of between-speaker scores for a given pair of speakers to be a scaled and shifted version of some base distribution defined by its \emph{cumulative distribution function} (CDF). Our earlier model \cite{Sholokhov2020-CSL} assumes a Gaussian base distribution. The following generalized algorithm allows to generate a set of between-speaker scores for given $N$: 
\begin{enumerate}
\item Sample $N$ pairs of location-scale values $\{(\mu_j, \sigma_j)\}_{j=1}^N$
\item Find the largest location parameter $\mu_k = \max_j \{ \mu_j \}$
\item Sample scores $\mathcal{S}_i=\{s_\ell\}_{\ell=1}^{L_i}$ by $s_\ell = \mu_k + \sigma_k F^{-1}(u_\ell)$, where $u_\ell \sim U[0,1]$ is uniformly-distributed.
\end{enumerate}
Here, $F(\cdot)$ is the CDF of the base distribution of scores. The algorithm uses \emph{inverse transform sampling} \cite{Devroye1986-book} to generate scores from the underlying distribution. Each pair $(\mu_j, \sigma_j)$ parameterizes the distribution of scores between a fixed target speaker and the $j$th impostor. It also assumes that the closest impostor has the largest location parameter $\mu_j$. One limitation of the model in \cite{Sholokhov2020-CSL} is the unrealistic assumption of Gaussian between-speaker scores. Here $F(\cdot)$ is allowed to be arbitrary. In practice, we use \texttt{torchpwl}\footnote{\url{https://pypi.org/project/torchpwl/}} to define a piece-wise linear function with monotonicity constraint for CDF approximation.

\subsection{PLDA-based model}

The above location-scale family of models represent speakers indirectly through their \emph{relative similarities} defined through between-speaker score distributions. The model described in this section uses, instead, latent identity variables to  represent individual speakers \emph{explicitly}. This gives an alternative predictor of $\mathrm{P}^N_\text{FA}(\tau)$ based on PLDA. 

A PLDA model with known parameters $\vec{\theta}_\text{plda}$ can be used to generate LLR scores, as follows.

\begin{enumerate}
\item Sample a pair of enrollment and test latent identity variables $(\vec{y}_\text{e}, \vec{y}_\text{t})$ from the prior: $\vec{y}_\text{e} \sim \mathcal{N}(\vec{0},\mtx{B})$ and $\vec{y}_\text{t}=\vec{y}_\text{e}$ under 
$H_0$; or draw the second sample $\vec{y}_\text{t} \sim \mathcal{N}(\vec{0},\mtx{B})$ under $H_1$.
\item Sample a pair of enrollment and test feature vectors 
\\
$\vec{\phi}_\text{e} \sim \mathcal{N}(\vec{y}_\text{e},\mtx{W})$, $\vec{\phi}_\text{t} \sim \mathcal{N}(\vec{y}_\text{t},\mtx{W})$ 
conditioned on the latent identity variables from the first step.
\item Compute the LLR score as $s = \ell(\vec{\phi}_\text{e},\vec{\phi}_\text{t})$ using \eqref{eq:plda-llr-score}.
\end{enumerate}


Note that the two first steps are stochastic, while the LLR score is a deterministic function of the sampled pair of feature vectors and the PLDA model. Under the $H_1$ hypothesis, this generative procedure yields scores of zero-effort impostors.
Following
Algorithm \ref{algo:scores-n-best}, it can be extended to sample $N>1$ impostors for a given target speaker. 
To select the closest impostor, we use the LLR score \eqref{eq:plda-llr-score} as a similarity measure between speakers.
Given identity variable of the closest impostor, one can sample a set of speaker-pair specific scores by repeating steps 2 and 3 in the algorithm above.

We include a learnable monotonic warping function applied to the scores generated by the model to increase flexibility. 

\subsection{Training methods}


We now describe a method for training generative models introduced above. The training data is a set of sets of between-speaker scores produced by any ASV system (black-box).
Generally, there are at least three alternative approaches to construct a regressor for predicting $\mathrm{P}^N_\text{FA}(\tau)$, given $N$ and $\tau$. 
The first one is to use any standard general-purpose regression technique 
to match model predictions and empirical estimates of $\mathrm{P}^N_\text{FA}(\tau)$ computed with \eqref{eq:fa-mc-general} from the empirical scores. Despite the apparent simplicity and attractiveness of such approach, and due to the lack of task-specific constraints, such models are exposed to a greater risk of failure for large values of $N$ \cite{Martius2017-extrapolation}.

The second approach is to follow a two-stage strategy: First, train a generative model of scores and use it to generate nontarget scores 
following Algorithm \ref{algo:scores-n-best}. Next, use the generated sets of scores to estimate $\mathrm{P}^N_\text{FA}(\tau)$ 
using \eqref{eq:fa-mc-general}.
We used this approach in \cite{Sholokhov2020-CSL} where 
a location-scale model with Gaussian base distribution was trained to maximize model log-likelihood \cite{Bishop-book}. Different from \cite{Sholokhov2020-CSL}, the models proposed in this work are instances of \emph{implicit} generative models: they are specified through a forward stochastic procedure for data generation, 
but do not allow direct likelihood evaluation \cite{Mohamed2016-implicit, Goodfellow2017-tutorial}. Even if implicit generative models can be trained using plethora of methods different from ML estimation (see \cite{Mohamed2016-implicit} and \cite{Louppe2019-adv}), it is 
non-trivial to design a training algorithm for models whose training set is a \emph{set of sets} (see, \emph{e.g.} \cite{Li-2018-pointcloud}), as is the case here.

In the last approach, generative model is also included to the prediction pipeline but trained \emph{discriminatively} by comparing the model-based estimates of $\mathrm{P}^N_\text{FA}(\tau)$ against the corresponding empirical estimates (treated as ground-truth). 
The regressor is trained by minimizing the mean square error (MSE) between the empirical and model-based false alarm rates. 
To address lack of differentiability, we replace the unit step function in \eqref{eq:fa-mc-general} by the sigmoid function with scaled argument. Also, the argmax function which appears in PLDA score generating algorithm was replaced by its approximation computed as a weighted sum of the speaker identity variables where weights are softmax-normalized similarities to the target speaker. This is similar to a so-called soft-attention mechanism introduced in \cite{Bahdanau2014-NMT}.

In contrast to purely generative training aimed at approximating  distribution of scores, discriminative training optimizes directly the final regression target. Using a restricted class of regression functions, in turn, allows us to keep the extrapolated values within the range of reasonable expectations. 


The resulting objective function (MSE in our experiments) includes random sampling and can be viewed as a \emph{nested} Monte-Carlo estimate of the expected loss. Generally, such MC estimates are biased for any finite $T$ \cite{Rainforth-2018} but useful for training via stochastic optimization, provided that $T$ is sufficiently large ($100$-$1000$ in our experiments). We used Adam \cite{Kingma2014-adam} optimizer with mini-batches of size $20$ and learning rate $10^{-3}$ to train both models. 
\section{Experiments}

We closely follow the experimental setup of \cite{Sholokhov2020-CSL}. We combine Voxceleb1 \cite{nagrani2017voxceleb} and Voxceleb2 \cite{Chung18b} corpora to have a dataset with a 
large number of speakers and sufficient number of utterances per speaker needed for reliable estimation of $P_\text{FA}^N$. The resulting dataset has $7365$ speakers with more than 100 utterances from each speaker, on average. The data was divided into three disjoint sets with $5345$, $40$ and $2000$ speakers. The first  set was used to train the ASV systems. The second one is the standard Voxceleb1 evaluation protocol \cite{Chung2019-voxsrc}, used as a sanity check of our ASV systems (see \cite{Sholokhov2020-CSL} for details). The third set which contains $1000$ male and $1000$ female speakers was used to compute scores for training models for $P_\text{FA}^N$ extrapolation. 
We computed similarity scores for each unique speaker pair of the same gender. To this end, we randomly selected 18 utterances for each of $2000$ speakers to obtain at least three hundreds of scores ($18^2=324$), which we assume to be sufficient to represent speaker-pair specific score distributions.

We used two standard ASV systems based on i-vectors and x-vectors to compute ASV scores used in our experiments. Due to the space limitations, we present results only for the x-vector system, which has EER of $3.61\%$ on the standard Voxceleb1 evaluation protocol. The key conclusions, however, are similar for the i-vector system. For more details on ASV systems and setup, refer to \cite{Sholokhov2020-CSL}.

\begin{figure}[!h]
\centering
\includegraphics[trim={0.6cm, 0.7cm, 0.0cm, 0.4cm},clip,height=7.0cm,width=7.8cm]{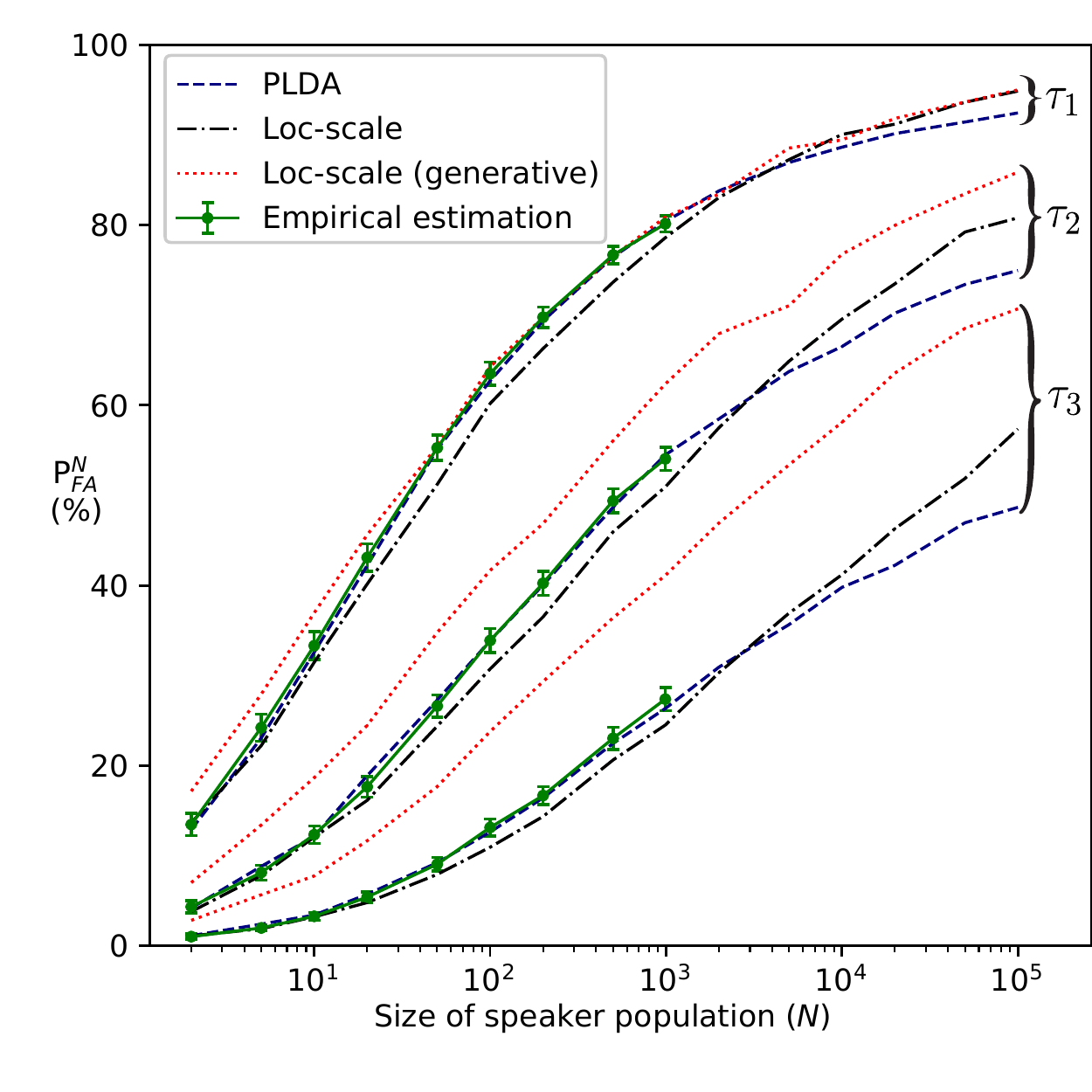}
\caption{Worst-case false alarm estimates for male scores given by x-vector system. Two new models and the model from \cite{Sholokhov2020-CSL} are shown along with the empirical estimates. 
The estimates are shown together with their 99\% confidence intervals.
}
\label{fig:extrapolation-all}
\end{figure}

We computed empirical and model-based estimates of the worst-case false alarm rates with $N$ impostors, $P_\text{FA}^N$, by randomly selecting a target speaker $T=1000$ times in Algorithm~\ref{algo:scores-n-best}. Fig. \ref{fig:extrapolation-all} shows the estimates obtained with different models. The three groups of curves correspond to different choices of ASV threshold, $\tau$. As detailed in \cite{Sholokhov2020-CSL}, these thresholds are the minimizers of three different detection cost functions (DCFs). The first DCF has high cost for misses ($\tau_1$), the second DCF has equal costs for misses and false alarms ($\tau_2$), and the last one penalizes false alarms more ($\tau_3$). The empirical curves end up to $N=1000$ impostors (as we have exhausted all data) while the 
extrapolated regression curves for $N > 1000$ may be used to speculate about the range of values of $P_\text{FA}^N$ for large sizes of impostor population. For instance, 
the ASV system with $P_\text{FA}^1 = P_\text{FA} \approx 1 \%$ may have the worst case false alarm rate around $50 \%$ for $N=10^5$. That is, if the attacker has a speech sample of the target speaker \emph{and} access to a proxy ASV system with comparable accuracy to the attacked one, the chance of accepting the closest impostor may reach $50\%$ for a population of $10^5$ available impostors.

To objectively assess the quality of models' forecasts, we measure mean absolute error (MAE) on the extrapolated values of $P_\text{FA}^N$ for a held-out set with $N\in[660, 999]$ while the corresponding empirical values (treated as the ground-truth and computed according to \eqref{eq:fa-mc-general}) were unseen by the model during training. In specific, the inputs in the training data were formed as pairs $(N, \tau)$ uniformly sampled from $[1, 660) \times [\tau_\text{min}, \tau_\text{max}]$, where the range of thresholds is determined according to the range of empirical scores. 
The held-out set was created similarly but with a different range of $N$.
The results 
summarised in Table \ref{tab:mae-all-models} 
indicate that more flexible models produce more accurate predictions. For instance, using a learnable base distribution instead of Gaussian decreases MAE for location-scale models and both models benefit from score warping. 
The location-scale and PLDA models have comparable accuracy. Importantly, both provide substantial improvement over earlier, purely generative model \cite{Sholokhov2020-CSL}.





\begin{table}[!h]
\centering
\caption{Extrapolation performance for different models in terms of MAE computed on the held-out set. $+$ indicates that a learnable score warping function was included to a model. We found that PLDA with $10$-dimensional feature space produces the best results.}
\begin{tabular}{|c|c|}
\cline{1-2}
Model                                & MAE, $\%$  \\ \cline{1-2}
Location-scale (Gaussian), generative \cite{Sholokhov2020-CSL} & 8.34 \\ 
Location-scale (Gaussian)    & 1.34 \\ 
Location-scale (Gaussian, $+$)       & 0.57 \\ \cline{1-2}
Location-scale (general CDF) & 0.67 \\ 
Location-scale (general CDF, $+$)    & 0.48 \\ \cline{1-2}
PLDA ($D=10$)                & 1.18 \\ 
PLDA ($D=10$, $+$)                   & 0.39 \\ \cline{1-2}
\end{tabular}
\label{tab:mae-all-models}
\end{table}

\section{Conclusions}

We advanced our recent work \cite{Vestman2020_mimicry_attack,Sholokhov2020-CSL} on worst-case impostors in the context of ASV. In specific, we introduced new tools for performance extrapolation of ASV systems. The models operate on detection score space and are therefore applicable outside the scope of ASV too. Our results indicate substantial improvement over our previous model \cite{Sholokhov2020-CSL}.

In future work, we may relax our worst-case impostor assumption, for instance so that the attacker fails to identify the closest impostor.
More generally, the usual assumption in adversarial machine learning where the attacker knows everything of the attacked system is potentially overly-pessimistic.   



\section{Acknowledgements}

This work was supported in part by the Academy of Finland (Proj. No. 309629).



\bibliographystyle{IEEEtran}
\bibliography{mybib}

\end{document}